\documentclass[%prb,%
twocolumn,
superscriptaddress,
footinbib,
amsmath,amssymb,
aps,
prl,
longbibliography
]{revtex4-1}
\usepackage[utf8]{inputenc} 
\usepackage[T1]{fontenc}
\usepackage{graphicx}% Include figure files
\usepackage{dcolumn}% Align table columns on decimal point
\usepackage{bm}% bold math
\usepackage{subcaption}
\usepackage{caption}
\usepackage{float}
\usepackage{lipsum}
\usepackage{amsmath}
\usepackage{amssymb}
\usepackage{dsfont}
\usepackage{color}
\usepackage{nccmath}
\usepackage[hidelinks,urlcolor=blue]{hyperref} 
\usepackage{xcolor}
\hypersetup{
	colorlinks,
	linkcolor={red!50!black},
	citecolor={blue},
	urlcolor={blue!80!black}
}
\usepackage[,%Uncomment any one of the following lines to test 
text={7.3in,10in},%centering,
]{geometry}
		{\end{pmatrix}\end{scriptsize}}%

\newcommand{\norm}[1]{\left\Vert#1\right\Vert}

\begin{document}
	
	%\preprint{APS/123-QED}
	
	\title{Local Marker for Interacting Topological Insulators}% Force line breaks with \\
	
	\author{A.A. Markov}
	\affiliation{
		Russian Quantum Center, Skolkovo IC, Bolshoy Bulvar 30, bld. 1, Moscow 121205, Russia 
	}
	\affiliation{Department of Physics, Lomonosov Moscow State University, Leninskie gory 1, 119991 Moscow, Russia}
	
	\author{A.N. Rubtsov}
	\affiliation{
		Russian Quantum Center, Skolkovo IC, Bolshoy Bulvar 30, bld. 1, Moscow 121205, Russia 
	}
	\affiliation{Department of Physics, Lomonosov Moscow State University, Leninskie gory 1, 119991 Moscow, Russia}

	\date{\today}% It is always \today, today,
	
	\begin{abstract}
  Topological states of matter were first introduced for non-interacting fermions on an infinite uniform lattice. Since then, substantial effort has been made to generalize these concepts to more complex settings. Recently, local markers have been developed that can describe the topological state of systems without translational symmetry and well-defined gap. However, no local marker for interacting matter has been proposed yet that is capable of directly addressing an interacting system. Here we suggest such a many-body local marker based on the single-particle Green's function. Using this marker we identify topological transitions in finite lattices of a Chern insulator with Anderson disorder and Hubbard interactions. Importantly, our proposal can be straightforwardly generalised to non-equilibrium systems.   	
  %		Local topological markers appear to be a powerful tool in the studies of topological matter allowing to treat inhomogeneous systems. To the best of our knowledge  In this letter we suggest such a many-body local marker based on the single-particle Green's functions and calculated it for a cluster of a Chern insulator with the Hubbard interactions. Importantly, our proposal can be generalised to the non-equilibrium settings in contrast with the topological Hamiltonian based approaches to the problem.   	
	\end{abstract}
	
	% Classification Scheme.Udep
	\maketitle
	\section{Introduction}
	
	The notion that gapped quantum many-body systems can be characterised by global topological invariants was first proposed by Thouless et al. \cite{thouless1982quantized}. However, the original construction was limited in scope, as it was only valid for non-interacting electrons on a translationally invariant lattice. Subsequent works have generalized the methods of momentum-space topology to interacting systems \cite{ishikawa1986magnetic,volovik2003universe,volovik2007quantum}. In addition, several methods have been used to extend these ideas to inhomogeneous systems, such as using twisted boundary conditions \cite{niu1985quantized,avron1985quantization} and non-commutative geometry \cite{bellissard1994noncommutative}. However, these methods only work on an infinite system with a band or mobility gap.

    Recently, a variety of quasi-topological characteristics, collectively known as local topological markers, have been developed \cite{kitaev2006anyons,bianco2011mapping,loring2011disordered}. Such markers characterise the topological properties of a system in real space, and thus are applicable to inhomogeneous systems with an ill-defined Brillouin Zone (BZ). Furthermore, having a quasi-local nature, they can be used for systems without a well-defined gap, e.g. for an interface between insulators with parameters corresponding to a different topological numbers \cite{bianco2014chern}. Such markers have been used to characterise topological phases in confined \cite{gebert2020local}, disordered \cite{loring2011disordered}, quasi-crystalline \cite{huang2018quantum,tran2015topological}, amorphous \cite{agarwala2019topological} and driven systems \cite{caio2019topological}.

    Several different local markers have been developed, most notably the Bott index \cite{loring2011disordered} and Chern marker \cite{kitaev2006anyons,bianco2011mapping}. These appear mathematically different, however the markers have much in common. Firstly,  they can be physically understood as the spatially-resolved Berry phase acquired by the system under magnetic flux insertion. They converge to the same value in the bulk of a system in thermodynamic limit \cite{toniolo2017equivalence}. Secondly, somewhat informally, the value of a marker answers the question: "What topological number would a system have if its neighborhood was repeated infinitely?". Therefore, the marker only has strict topological meaning when averaged over an infinite number of sites. Finally, all of them are constructed using a projector onto the filled single-particle states, preventing straightforward generalisations of this concept to the interacting systems.

	Recent attempts to extend the notion of local topological markers to interacting system include \cite{amaricci2017edge,irsigler2019interacting}. To this end, it was proposed to use an effective non-interacting topological Hamiltonian \cite{wang2012simplified}. To the best of our knowledge, the only local topological markers capable of addressing a many-body system rely on connecting it to a related non-interacting system first. No local marker exists that can directly address an interacting system. Additionally, the definition of a topological Hamiltonian relies on the notion of Matsubara freuqncies, so any approach that uses this method will not work for systems out of equilibrium.
	
	In this letter we propose a local topological marker for interacting systems. It is expressed in terms of the single particle Green's functions, so is suitable for numerical calculations with standard many-body techniques, e.g. exact diagonalisation of small clusters or schemes \cite{potthoff1999metallic,takemori2018intersite} based on dynamical mean-field theory \cite{georges1996dynamical} and its extensions \cite{rubtsov2008dual}. Our marker is applicable for systems without symmetries - unitary class A in terms of the ten-fold classification \cite{altland1997nonstandard}, but the generalization to the other symmetry-protected topological phases seems to be possible following the lines of ref. \cite{wang2010topological}.
	
	The manuscript is organized as follows: in the first section we introduce our marker and review some of its main properties. Thereafter, we present a study of the marker's behaviour in disordered and interacting systems, solved using exact diagonalisation. We conclude with a comparison of our marker against the topological Hamiltonian approach and propose possible extensions of the present study.

	\section{Local Green marker}
	\label{sec:LGM}
	 \begin{figure}[t!]
		\centering
		\includegraphics[width=1.0\linewidth]{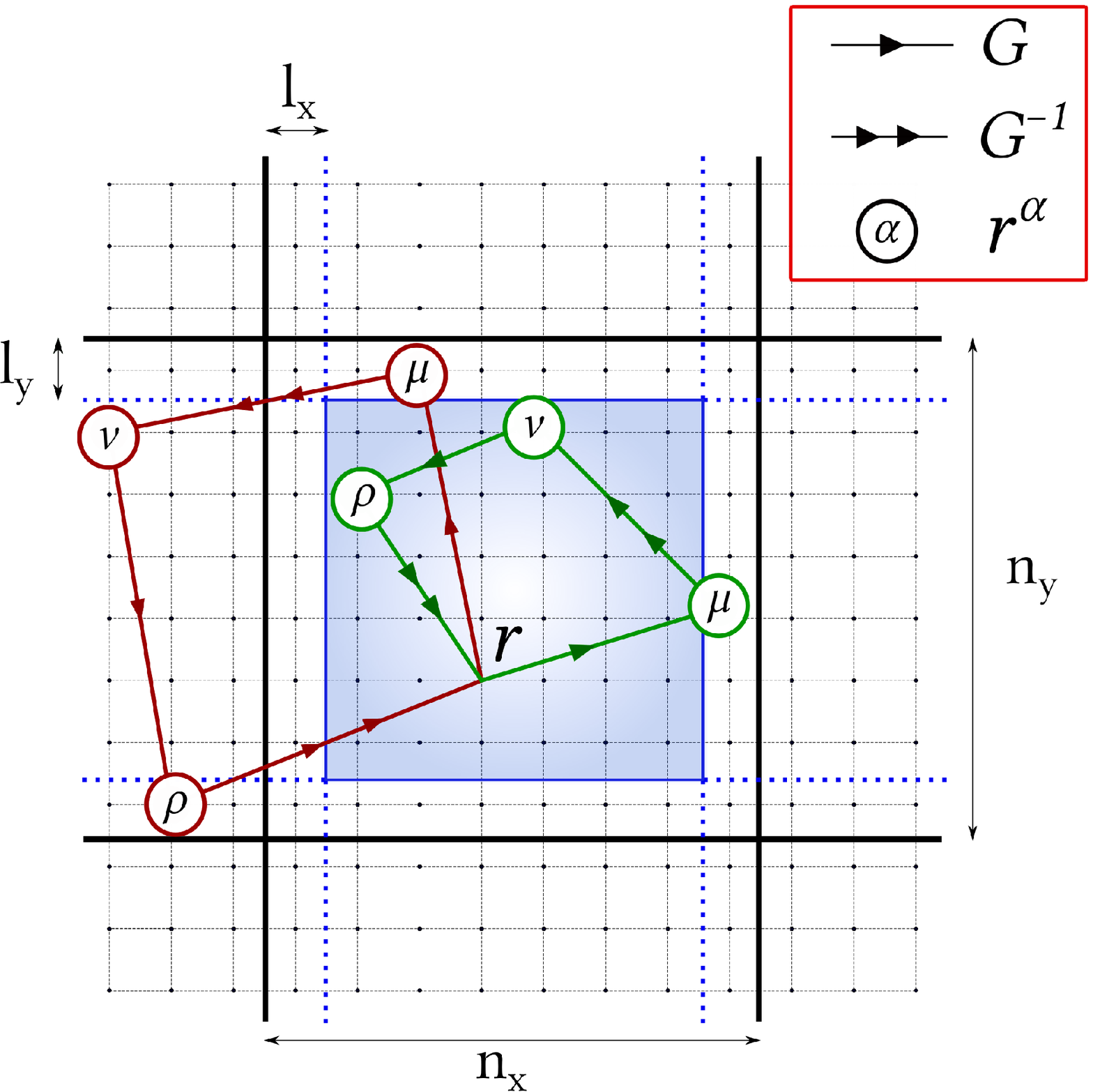}
		\caption{Diagrammatic representation of the formula for the local Green marker and the geometry of the problem. The black lines depicts the borders of a unit supercell. The blue region is the ''bulk'' of the system, separated from the borders by $l_x$ and $l_y$ sites (in the particular case depicted $l_x = l_y = 1$). The LGM at a given point $\bm{r}$ is a sum of terms represented by all possible loops originating at $\bm{r}$ and traversing the real space. In the thermodynamic limit, only loops traversing a single supercell - the green ones on the picture gives non-vanishing contribution in the bulk. While all other loops (red ones) contribution passes to zero. }
		\label{fig:LGMdiagram}
	\end{figure}
	
	In a seminal work \cite{ishikawa1986magnetic}, Ishikawa and Matsuyama managed to express the Hall conductivity of an interacting translationally invariant system with a unique ground state in terms of the single-particle Greens functions only:
	\begin{equation}
	C = \frac{1}{24 \pi^2} \epsilon_{\alpha\beta \gamma}\int d^3k\; Tr(G\partial_{k_\alpha}G^{-1}G\partial_{k_\beta}G^{-1}G\partial_{k_\gamma}G^{-1}).
	\label{eq:green_winding_number}
	\end{equation}
	Here, $\epsilon_{\alpha\beta \gamma}$ is the totally-antisymmetric tensor, $G$ is the Matsubara Green's function and the trace is taken over internal (e.g. orbital or spin) degrees of freedom. The convention $k_0\widehat{=}i\omega$ ($i\omega$ is continuous Matsubara frequency) was used in order to obtain the symmetric form with respect to frequency and momentum variables. The expression in eqn. (\ref{eq:green_winding_number}) can be understood as a topological invariant of the map $G(i\omega,k): T^3 \to GL(N,\mathbb{C})$ from momentum-frequency space to the space of invertible matrices. Loosely speaking, it counts the number of times $T^3$ wraps around the single $S^3$ skeleton of the $U(N)$ CW-complex. 
	
	Our construction of a local marker starts with the generalised Ishikawa-Matsuyama invariant, proposed in ref. \cite{zheng2019interaction}. In a disordered system, the Hall conductivity may be approximated by tiling the system, treating it as a supercell in a larger superlattice. This allows us to calculate the Hall conductivity in a form of a winding number:
	
	\begin{equation}
	\begin{split}
	\sigma_{xy} = \frac{\epsilon^{\mu\nu \rho}}{12 \pi N} \int d\omega \;\sum_{\bm{\theta}}
	Tr_{s}\left(G^{\bm{\theta}}\partial_{\mu}\left[G^{\bm{\theta}}\right]^{-1}\right.\\
	\left.\times
	G^{\bm{\theta}}\partial_{\nu}\left[G^{\bm{\theta}}\right]^{-1}G^{\bm{\theta}}\partial_{\rho}\left[G^{\bm{\theta}}\right]^{-1}\right),
	\label{eq:GIMP}
	\end{split}
	\end{equation}	
	where the trace $Tr_{s}$ is taken over all supercell degrees of freedom and  $G^{\bm{\theta}}$ is the Fourier transform of the Matsubara Green's function:
	
	\begin{equation}
	G_{i\omega}^{\bm{\theta}}(\bm{r},\bm{r'}) =  \sum_{\bm{R}} G(\bm{R}+\bm{r},\bm{r'})e^{i\bm{\theta}(\bm{R}+\bm{r} - \bm{r'})}.
	\label{eq:EISGreen}
	\end{equation}
	
	Here, $\bm{R}$ denotes the position of a supercell origin and  $\bm{r}$ specifies the site in a given supercell. 
	
	Our goal is to express eqn.~\ref{eq:GIMP} as an average of a real space quantity. It has been shown that the winding number of a function can be calculated from its Fourier transformed coefficients provided that the function has bounded mean oscillation \cite{brezis1995degree}. That means we can safely return to real space in order to calculate the invariant (eqn. \ref{eq:GIMP}). Then, the winding number may be calculated as a supercell bulk average, up to corrections vanishing as $n_{x}$ and $n_y$ go to infinity (see supplemental material \footnote{See supplemental material at [URL will be inserted by publisher] for the details of the derivation of the local Green marker. It includes ref.~ \cite{ceresoli2007orbital,jaffard1990proprietes,benzi2016localization} absent in the main text.} for the details): 
	
	 \begin{equation}
       \sigma_{xy} = \frac{1}{N_{bulk}} \sum_{\bm{r}\in bulk}\mathfrak{G}(\bm{r}),
       \label{eq:average}
   \end{equation}	
   where $N_{bulk}$ is the number of sites in the bulk of a supercell. 

		\begin{equation}
		\begin{split}
       \mathfrak{G}(\bm{r})=-i \frac{\epsilon_{\mu\nu \rho}}{6}  \sum_{\bm{r}_i} Tr \left( \vphantom{ \sum_{r_i} } G(\bm{r},\bm{r}_1)\bm{r}_1^{\mu} G^{-1}(\bm{r}_1,\bm{r}_2)\bm{r}_2^\nu \right. \\ \left. \times   G(\bm{r}_2,\bm{r}_3)\bm{r}_3^\rho G^{-1}(\bm{r}_3,\bm{r})\vphantom{ \sum_{r_i} } \right).
       \end{split}
       \label{eq:LGM1}
  \end{equation}
	Here, $\bm{r}_0$ refers to imaginary time and  the trace is taken over spin-orbital degrees of freedom. This formula admits a graphical representation reminiscent of Feynman diagrams, presented in fig.~\ref{fig:LGMdiagram}. Each term in eqn. \ref{eq:LGM1} can be represented as a ``loop'' in a real space-time originating from a given point $\bm{r}$. Green's functions are represented by arrows connecting two points, their inverses by double arrows and factors of $\bm{r}^{\alpha}$ in (\ref{eq:LGM1}) as vertices with labels which telling you space-time component should be taken. 
	
	This is our suggestion for the local marker, which we call local Green marker (LGM). In an extended bulk the LGM evaluates the known topological invariant. It has a quasi-local nature due to the exponential decay of the correlators with distance in a gapped phase with decaying interactions \cite{hastings2006spectral,nachtergaele2006lieb,wang2020tightening}. Also, one can demonstrate, using  $Tr(A) = Tr(A^T)$, that the average of LGM over a finite lattice vanishes, as does the average of the local Chern maker \cite{kitaev2006anyons, bianco2011mapping}. 
	
	Intuitively, the topological meaning of LGM in an infinite disordered system can be understood by reversing the logic of our derivation. At a given site $\bm{i}$,  $\mathfrak{G}(\bm{i})$ approximates the value of the invariant (\ref{eq:GIMP}) for a crystalline system consisting of supercells built from the neighborhood of the site $\bm{i}$.

	\section{Numerical Results}
	
    As a proof of concept we numerically studied a Chern insulator with phase transitions driven by Anderson disorder and Hubbard on-site interactions. We chose the Qi-Wu-Zhang (QWZ) model proposed in \cite{qi2006topological}. In real space its Hamiltonian can be written as:
	
	\begin{align}
	\label{eq:QWZ}
	\hat{H}_0 =  t \sum_{i}&\left( \frac{(\sigma_z + i\sigma_x)^{\alpha\beta}}{2}c^{\dagger}_{i+1_x,\alpha} c_{i,\beta}+h.c.\right)\nonumber\\
	+t &\sum_{i}\left( \frac{(\sigma_z +i\sigma_y)^{\alpha\beta}}{2}c^{\dagger}_{i+1_y,\alpha} c_{i,\beta}+h.c.\right)\\
	+ &\sum_{i}\left( u \sigma_z^{\alpha\beta}+ \mu \sigma_0^{\alpha\beta}\right)c^{\dagger}_{i,\alpha} c_{i,\beta}\nonumber,
	\end{align}
	where the Pauli matrices act in the two-level spin subspace and $\sigma_0\equiv \mathds{1} $ is the identity matrix. The model is in a non-trivial topological insulating phase when the on-site magnetic field $u$ is in the range  $(-2,2) / \{0\}$. From now on we set $t=1$, fixing the energy scale and choose $u$ equal to $-1$, so $H_0$ corresponds to a topological phase with $\sigma_{xy} = \frac{1}{2\pi} $.
	
	First, we study a phase transition caused by uniformly distributed Anderson disorder:
	
	\begin{equation}
	\begin{split}
	\hat{H}_A &= \sum_{i}\varepsilon_i c^{\dagger}_{i,\alpha} c_{i,\alpha}, \qquad  P_A(\varepsilon) = \frac{1}{W}\theta\left(\frac{W}{2}-|\varepsilon| \right)\\
	\hat{H} &= \hat{H}_0 + \hat{H}_A.
     \end{split}
	\end{equation}
    Here $P_A(\varepsilon)$ is the probability density for the on-site random potential $\varepsilon$.
    
\begin{figure}[t!]
		\centering
		\includegraphics[width=0.9\linewidth]{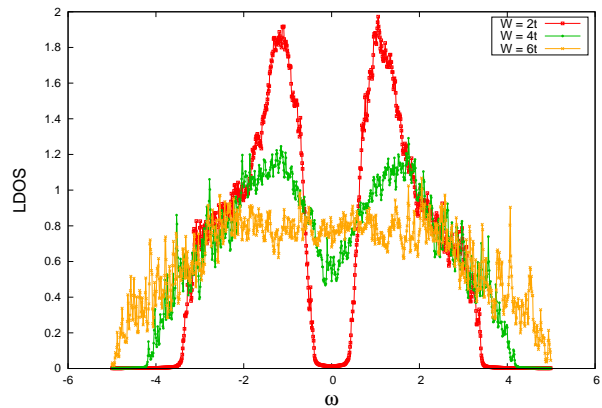}
		\caption{Local density of states $A_{\bm{r}}(\omega) = -\frac{1}{\pi}Im(G(\bm{r},\bm{r};\omega+i\delta))$ at the middle of a finite $10\times10$ sample averaged over $500$ realisation of Anderson disorder for different values of disorder strength $W$.}
		\label{fig:AndLDOS}
	\end{figure}

    The transition between topological band insulator and localized Anderson insulator happens at a finite disorder strength $W_c$, in contrast to the metal-insulator transition, which is absent in 2-d. This insulator to insulator transition was extensively studied in the literature \cite{bellissard1994noncommutative,onoda2007localization,castro2015anderson} and its mechanism established. As disorder strength grows, in-gap localized states gradually displace extended band states. At a critical disorder strength $W_c$ bands carrying non-trivial Chern numbers annihilate each other \cite{onoda2007localization}, this process is illustrated in fig. \ref{fig:AndLDOS}.
    
   	 \begin{figure}[t!]
		\centering
		\includegraphics[width=1.0\linewidth]{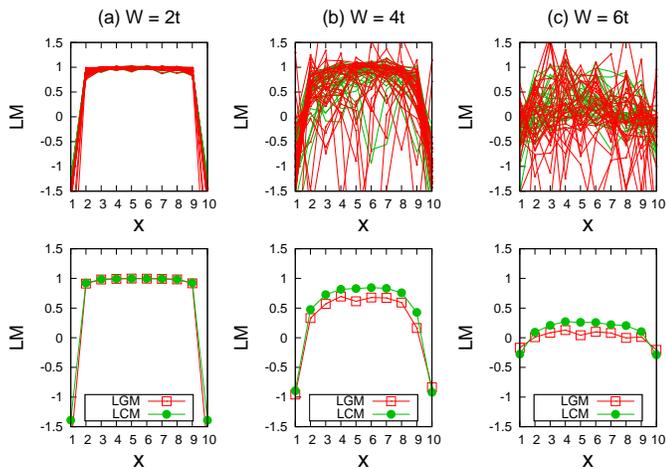}
		\caption{Values of the local Green and local Chern marker in the presence of the Anderson disorder. The first row demonstrates their values for different disorder instances, the second  - their averages over 500 realisations.}
		\label{fig:andLM}
	\end{figure}
	
	Bellisard \cite{bellissard1994noncommutative} demonstrated that the average over disorder of the local Chern marker from \cite{bianco2011mapping} is quantised to integer values in the thermodynamic limit. This result can already be seen from calculations of the local Chern marker for relatively small samples. The same can be concluded for the LGM as presented in fig. \ref{fig:andLM}. Typical values of the marker are distributed around $C = 1$, even at moderate disorder strength ($W=4t$) close to the critical value as indictes the fig. \ref{fig:andLM}(b). However, where conducting states are present the marker takes a large negative value. Thus, as in finite samples this value of disorder strength ($W=4t$) is large enough to create ocassionaly metallic-like extended states, the average values of the LGM and Chern marker are somewhat lower than expected, with a bulk mean around $C_{av} \approx 0.5$. When the disorder strength is strong enough (fig. \ref{fig:andLM}(c)), the material stops being topological and the LGM averages to zero.

	As a next step we considered the Chern insulator - Mott insulator transition driven by the Hubbard interactions:

	\begin{equation}
	\begin{split}
	\hat{H}_I &= \sum_{i} U \hat{n}_{i\downarrow} \hat{n}_{i\uparrow}\\
	\hat{H} &= \hat{H}_0 + \hat{H}_I.
     \end{split}
	\end{equation}
	
	We calculated the Green's functions using the Lanczos algorithm \cite{dagotto1994correlated} for a Hubbard cluster of $4\times4$  sites with open boundary conditions. We used the HPhi library as a solver \cite{kawamura2017quantum}. A single calculation of a Green's function $G_{i \omega}(\bm{r},\bm{r}')$ at given values of position and spin variables requires two iterations of the full Lanczos algorithm. This put an exact diagonalization calculation of the LGM for a $4\times4$ system at the edge of modern computational capacity, even for symmetric clusters. However, already at such small clusters one can see some of the most important features of the phase transition \cite{varney2010interaction}.      
	
\begin{figure}[t!]
		\centering
		\includegraphics[width=0.8\linewidth]{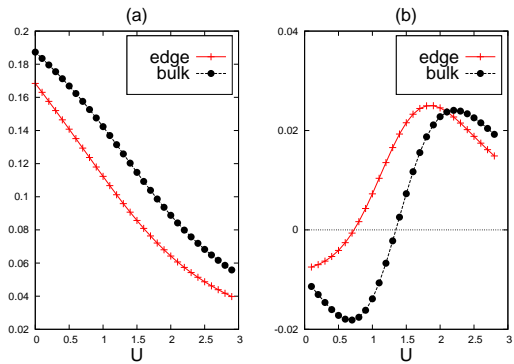}
		\caption{ Double occupancy (a) $\Delta_i$ and its second derivative (b) as functions of interaction strength $U$ calculated in the middle and at the edge of the cluster. Note that the transition happens at lower values of $U$ at the edge, compared to the bulk, since the kinetic energy term is weaker at the edges.}
		\label{fig:double}
	\end{figure}

	At finite sample size a true Mott phase transition does not occur. However, signs of the Mott transition can be found in the dependence of the double occupancy $\Delta_i = \langle \hat{n}_{i\uparrow} n_{i\downarrow} \rangle$ on interaction strength $U$ as well as from the behaviour of the local density of states at different values of $U$. The derivative of the double occupancy diverges at the critical point \cite{rozenberg1994metal}. In a finite sample this non-analytic behaviour is smoothed, however its footprint may be captured from the dependence of the second derivative of $\Delta_i$ as illustrated in fig. \ref{fig:double}. 
	
\begin{figure}[t!]
		\centering
		\includegraphics[width=1.0\linewidth]{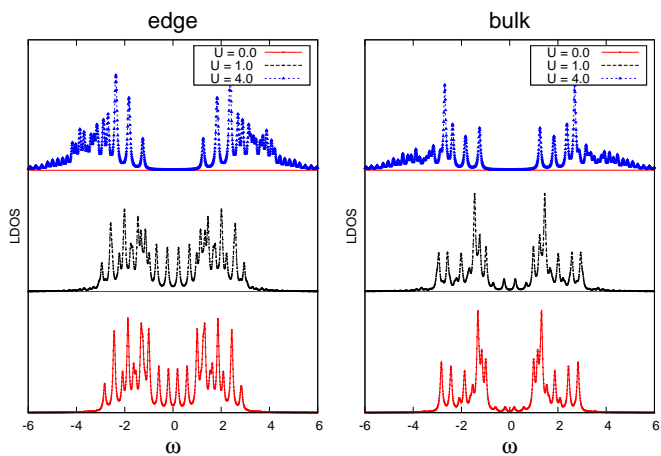}
		\caption{Local density of states for different values of $U$ at the edge and in the middle of the $4\times4$ interacting cluster.}
		\label{fig:LDOS}
	\end{figure}

	For open boundaries another important sign of the Mott transition is a disappearance of the edge modes. This characteristic feature is captured by the local density of states (fig. \ref{fig:LDOS}). 
	
\begin{figure}[t!]
		\centering
		\includegraphics[width=1.0\linewidth]{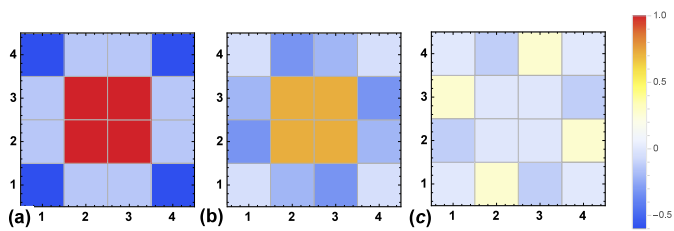}
		\caption{Local Green marker at different interaction strengths calculated for a Chern insulator $4\times4$ cluster with Hubbard onsite interactions.}
		\label{fig:LGM}
	\end{figure}

	The LGM turned to be a reasonable marker for this phase transition. At intermediate values of interaction strength (at $U=1$) in the middle of the transition region, its ``bulk'' value gets suppressed from the non-interacting value, see Fig. \ref{fig:LGM}. Deeper in the Mott phase (at $U=4$) the LGM averages to almost zero.  
	
	\section{Conclusions and Outlook}
	
	We proposed a topological marker for interacting systems that does not rely on reference to a non-interacting Hamiltonian. To the best of our knowledge, such a marker has not previously been proposed. We tested our marker in several contexts, calculating it exactly in the presence of the disorder and interactions.
	
	From a computational perspective, our invariant is harder to calculate than the local Chern marker obtained from the topological Hamiltonian, as was done in~\cite{amaricci2017edge,irsigler2019interacting}. Our marker involves integration over imaginary time (or summation over all the Matsubara frequencies in equilibrium), which imposes substantial numerical requirements.
	
	However, the proposal seems attractive for several reasons. Firstly, the local markers are not exactly topological invariants. They has a quasi-topological nature, so their local value is not invariant under smooth deformation of the system. Only the average of a marker over an infinite number of sites is topologically robust. Thus, an approach based on smoothly removing interactions, as is done in the topological Hamiltonian approach, and then calculating the LCM of the non-interacting system has not been shown to give us information that exactly corresponds to the interacting system. Secondly, our approach extends the applicability of topological markers to systems which cannot be adiabatically connected to a non-interacting topological Hamiltonian. An example is systems with a nontrivial frequency-domain winding number \cite{wang2011frequency}, although such exotic phases are of unproved relevance for the field of topological matter \cite{rachel2018interacting}. Lastly, our approach allows us to study Local topological markers in non-stationary settings, for example in quenched topological matter. In the case of adiabatic evolution, the use of a topological Hamiltonian can be put on solid grounds, however for fast dynamics it looks inapplicable. On the other hand, our approach can be performed using the Keldysh formalism \cite{keldysh1965diagram}. Additionally, it is compatible with the earlier proposals for calculating the winding number (\ref{eq:green_winding_number}) in interacting systems out of equilibrium \cite{foster2013quantum}. 
	
	Let us briefly discuss several possible future directions for investigation. Firstly, it would be worthwhile to compare numerical calculations of the local Green marker in larger non homogeneous systems to the topological Hamiltonian based approach in \cite{amaricci2017edge,irsigler2019interacting}. Secondly, an important question is to what extent our marker can be used for fractional Chern insulators. The applicability of the Ishikawa-Matsuyama invariant to such systems is an open and interesting question, see discussions in \cite{budich2013adiabatic,ishikawa1986magnetic}. Finally, we believe our marker to be very attractive to study in systems out of equilibrium.

	\section{Acknowledgments}
	
	We thank Peru d'Ornellas for many interesting and useful discussions and for critical reading, commenting and edditing of this manuscript. Useful discussions with Georg Rohringer are gratefully acknowledged. The work was supported by the North-German Supercomputing Alliance (HLRN) and was carried out in the framework of the Roadmap for Quantum computing in Russia. A.A.M. is also supported by the ``Basis'' foundation under grant $\#$18-3-01. 
	
	\bibliography{LGM}		

\begin{thebibliography}{10}

\bibitem{thouless1982quantized}
David~J Thouless, Mahito Kohmoto, M~Peter Nightingale, and Marcel den Nijs.
\newblock Quantized {H}all conductance in a two-dimensional periodic potential.
\newblock {\em Physical review letters}, 49(6):405, 1982.

\bibitem{ishikawa1986magnetic}
Kenzo Ishikawa and Toyoki Matsuyama.
\newblock Magnetic field induced multi-component {QED}3 and quantum {H}all
  effect.
\newblock {\em Zeitschrift f{\"u}r Physik C Particles and Fields},
  33(1):41--45, 1986.

\bibitem{volovik2003universe}
Grigory~E Volovik.
\newblock {\em The universe in a helium droplet}, volume 117.
\newblock Oxford University Press on Demand, 2003.

\bibitem{volovik2007quantum}
Grigory~E Volovik.
\newblock Quantum phase transitions from topology in momentum space.
\newblock In {\em Quantum analogues: from phase transitions to black holes and
  cosmology}, pages 31--73. Springer, 2007.

\bibitem{niu1985quantized}
Qian Niu, Ds~J Thouless, and Yong-Shi Wu.
\newblock Quantized {H}all conductance as a topological invariant.
\newblock {\em Physical Review B}, 31(6):3372, 1985.

\bibitem{avron1985quantization}
Joseph~E Avron and Ruedi Seiler.
\newblock Quantization of the {H}all conductance for general, multiparticle
  {S}chr{\"o}dinger {H}amiltonians.
\newblock {\em Physical review letters}, 54(4):259, 1985.

\bibitem{bellissard1994noncommutative}
Jean Bellissard, Andreas van Elst, and Hermann Schulz-Baldes.
\newblock The noncommutative geometry of the quantum {H}all effect.
\newblock {\em Journal of Mathematical Physics}, 35(10):5373--5451, 1994.

\bibitem{kitaev2006anyons}
Alexei Kitaev.
\newblock Anyons in an exactly solved model and beyond.
\newblock {\em Annals of Physics}, 321(1):2--111, 2006.

\bibitem{bianco2011mapping}
Raffaello Bianco and Raffaele Resta.
\newblock Mapping topological order in coordinate space.
\newblock {\em Physical Review B}, 84(24):241106, 2011.

\bibitem{loring2011disordered}
Terry~A Loring and Matthew~B Hastings.
\newblock Disordered topological insulators via {C}*-algebras.
\newblock {\em EPL (Europhysics Letters)}, 92(6):67004, 2011.

\bibitem{bianco2014chern}
Raffaello Bianco.
\newblock {C}hern invariant and orbital magnetization as local quantities.
\newblock 2014.

\bibitem{gebert2020local}
Urs Gebert, Bernhard Irsigler, and Walter Hofstetter.
\newblock Local {C}hern marker of smoothly confined {H}ofstadter fermions.
\newblock {\em Physical Review A}, 101(6):063606, 2020.

\bibitem{huang2018quantum}
Huaqing Huang and Feng Liu.
\newblock Quantum spin {H}all effect and spin bott index in a quasicrystal
  lattice.
\newblock {\em Physical review letters}, 121(12):126401, 2018.

\bibitem{tran2015topological}
Duc-Thanh Tran, Alexandre Dauphin, Nathan Goldman, and Pierre Gaspard.
\newblock Topological {H}ofstadter insulators in a two-dimensional
  quasicrystal.
\newblock {\em Physical Review B}, 91(8):085125, 2015.

\bibitem{agarwala2019topological}
Adhip Agarwala.
\newblock Topological insulators in amorphous systems.
\newblock In {\em Excursions in Ill-Condensed Quantum Matter}, pages 61--79.
  Springer, 2019.

\bibitem{caio2019topological}
Marcello~Davide Caio, Gunnar M{\"o}ller, Nigel~R Cooper, and MJ~Bhaseen.
\newblock Topological marker currents in {C}hern insulators.
\newblock {\em Nature Physics}, 15(3):257--261, 2019.

\bibitem{toniolo2017equivalence}
Daniele Toniolo.
\newblock On the equivalence of the {B}ott index and the {C}hern number on a
  torus, and the quantization of the {H}all conductivity with a real space kubo
  formula.
\newblock {\em arXiv preprint arXiv:1708.05912}, 2017.

\bibitem{amaricci2017edge}
A~Amaricci, L~Privitera, F~Petocchi, M~Capone, G~Sangiovanni, and B~Trauzettel.
\newblock Edge state reconstruction from strong correlations in quantum spin
  {H}all insulators.
\newblock {\em Physical Review B}, 95(20):205120, 2017.

\bibitem{irsigler2019interacting}
Bernhard Irsigler, Jun-Hui Zheng, and Walter Hofstetter.
\newblock Interacting {H}ofstadter interface.
\newblock {\em Physical review letters}, 122(1):010406, 2019.

\bibitem{wang2012simplified}
Zhong Wang and Shou-Cheng Zhang.
\newblock Simplified topological invariants for interacting insulators.
\newblock {\em Physical Review X}, 2(3):031008, 2012.

\bibitem{potthoff1999metallic}
M~Potthoff and W~Nolting.
\newblock Metallic surface of a {M}ott insulator--{M}ott insulating surface of
  a metal.
\newblock {\em Physical Review B}, 60(11):7834, 1999.

\bibitem{takemori2018intersite}
Nayuta Takemori, Akihisa Koga, and Hartmut Hafermann.
\newblock Intersite electron correlations on inhomogeneous lattices: a
  real-space dual fermion approach.
\newblock {\em arXiv preprint arXiv:1801.02441}, 2018.

\bibitem{georges1996dynamical}
Antoine Georges, Gabriel Kotliar, Werner Krauth, and Marcelo~J Rozenberg.
\newblock Dynamical mean-field theory of strongly correlated fermion systems
  and the limit of infinite dimensions.
\newblock {\em Reviews of Modern Physics}, 68(1):13, 1996.

\bibitem{rubtsov2008dual}
AN~Rubtsov, MI~Katsnelson, and AI~Lichtenstein.
\newblock Dual fermion approach to nonlocal correlations in the {H}ubbard
  model.
\newblock {\em Physical Review B}, 77(3):033101, 2008.

\bibitem{altland1997nonstandard}
Alexander Altland and Martin~R Zirnbauer.
\newblock Nonstandard symmetry classes in mesoscopic normal-superconducting
  hybrid structures.
\newblock {\em Physical Review B}, 55(2):1142, 1997.

\bibitem{wang2010topological}
Zhong Wang, Xiao-Liang Qi, and Shou-Cheng Zhang.
\newblock Topological order parameters for interacting topological insulators.
\newblock {\em Physical review letters}, 105(25):256803, 2010.

\bibitem{zheng2019interaction}
Jun-Hui Zheng, Tao Qin, and Walter Hofstetter.
\newblock Interaction-enhanced integer quantum {H}all effect in disordered
  systems.
\newblock {\em Physical Review B}, 99(12):125138, 2019.

\bibitem{brezis1995degree}
Ha{\i}m Brezis and Louis Nirenberg.
\newblock Degree theory and {BMO}; part i: Compact manifolds without
  boundaries.
\newblock {\em Selecta Mathematica New Series}, 1(2):197--264, 1995.

\bibitem{Note1}
See supplemental material at [URL will be inserted by publisher] for the
  details of the derivation of the local Green marker. It includes ref.~ \cite
  {ceresoli2007orbital,jaffard1990proprietes,benzi2016localization} absent in
  the main text.

\bibitem{hastings2006spectral}
Matthew~B Hastings and Tohru Koma.
\newblock Spectral gap and exponential decay of correlations.
\newblock {\em Communications in mathematical physics}, 265(3):781--804, 2006.

\bibitem{nachtergaele2006lieb}
Bruno Nachtergaele and Robert Sims.
\newblock Lieb-robinson bounds and the exponential clustering theorem.
\newblock {\em Communications in mathematical physics}, 265(1):119--130, 2006.

\bibitem{wang2020tightening}
Zhiyuan Wang and Kaden~RA Hazzard.
\newblock Tightening the lieb-robinson bound in locally interacting systems.
\newblock {\em PRX Quantum}, 1(1):010303, 2020.

\bibitem{qi2006topological}
Xiao-Liang Qi, Yong-Shi Wu, and Shou-Cheng Zhang.
\newblock Topological quantization of the spin {H}all effect in two-dimensional
  paramagnetic semiconductors.
\newblock {\em Physical Review B}, 74(8):085308, 2006.

\bibitem{onoda2007localization}
Masaru Onoda, Yshai Avishai, and Naoto Nagaosa.
\newblock Localization in a quantum spin {H}all system.
\newblock {\em Physical review letters}, 98(7):076802, 2007.

\bibitem{castro2015anderson}
Eduardo~V Castro, M~Pilar L{\'o}pez-Sancho, and Mar{\'\i}a~AH Vozmediano.
\newblock Anderson localization and topological transition in {C}hern
  insulators.
\newblock {\em Physical Review B}, 92(8):085410, 2015.

\bibitem{dagotto1994correlated}
Elbio Dagotto.
\newblock Correlated electrons in high-temperature superconductors.
\newblock {\em Reviews of Modern Physics}, 66(3):763, 1994.

\bibitem{kawamura2017quantum}
Mitsuaki Kawamura, Kazuyoshi Yoshimi, Takahiro Misawa, Youhei Yamaji, Synge
  Todo, and Naoki Kawashima.
\newblock Quantum lattice model solver {H}$\phi$.
\newblock {\em Computer Physics Communications}, 217:180--192, 2017.

\bibitem{varney2010interaction}
Christopher~N Varney, Kai Sun, Marcos Rigol, and Victor Galitski.
\newblock Interaction effects and quantum phase transitions in topological
  insulators.
\newblock {\em Physical Review B}, 82(11):115125, 2010.

\bibitem{rozenberg1994metal}
Marcelo~J Rozenberg, Goetz Moeller, and Gabriel Kotliar.
\newblock The metal--insulator transition in the {H}ubbard model at zero
  temperature ii.
\newblock {\em Modern Physics Letters B}, 8(08n09):535--543, 1994.

\bibitem{wang2011frequency}
Lei Wang, Xi~Dai, and XC~Xie.
\newblock Frequency domain winding number and interaction effect on topological
  insulators.
\newblock {\em Physical Review B}, 84(20):205116, 2011.

\bibitem{rachel2018interacting}
Stephan Rachel.
\newblock Interacting topological insulators: a review.
\newblock {\em Reports on Progress in Physics}, 81(11):116501, 2018.

\bibitem{keldysh1965diagram}
Leonid~V Keldysh et~al.
\newblock Diagram technique for nonequilibrium processes.
\newblock {\em Sov. Phys. JETP}, 20(4):1018--1026, 1965.

\bibitem{foster2013quantum}
Matthew~S Foster, Maxim Dzero, Victor Gurarie, and Emil~A Yuzbashyan.
\newblock Quantum quench in a p+ip superfluid: Winding numbers and topological
  states far from equilibrium.
\newblock {\em Physical Review B}, 88(10):104511, 2013.

\bibitem{budich2013adiabatic}
Jan~Carl Budich and Bj{\"o}rn Trauzettel.
\newblock From the adiabatic theorem of quantum mechanics to topological states
  of matter.
\newblock {\em physica status solidi (RRL)--Rapid Research Letters},
  7(1-2):109--129, 2013.

\bibitem{ceresoli2007orbital}
Davide Ceresoli and Raffaele Resta.
\newblock Orbital magnetization and {C}hern number in a supercell framework:
  {S}ingle k-point formula.
\newblock {\em Physical Review B}, 76(1):012405, 2007.

\bibitem{jaffard1990proprietes}
Stephane Jaffard.
\newblock Propri{\'e}t{\'e}s des matrices {\guillemotleft}bien
  localis{\'e}es{\guillemotright} pr{\`e}s de leur diagonale et quelques
  applications.
\newblock In {\em Annales de l'Institut Henri Poincare (C) Non Linear
  Analysis}, volume~7, pages 461--476. Elsevier, 1990.

\bibitem{benzi2016localization}
Michele Benzi.
\newblock Localization in matrix computations: Theory and applications.
\newblock In {\em Exploiting Hidden Structure in Matrix Computations:
  Algorithms and Applications}, pages 211--317. Springer, 2016.

\end{thebibliography}
	\begin{widetext}
	\section{Supplemental Material}
		\begin{figure}[H]
		\centering
		\includegraphics[width=0.5\linewidth]{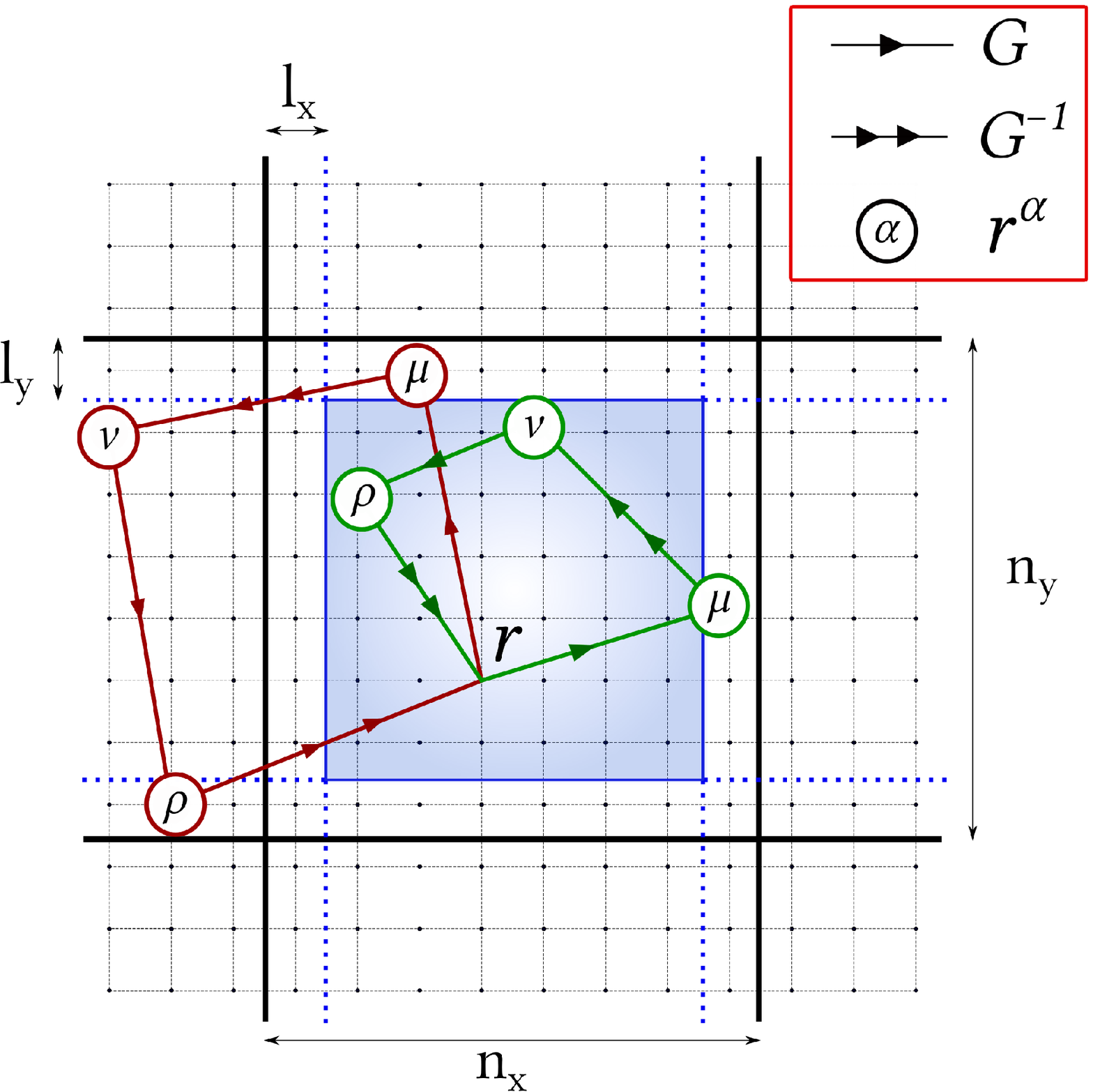}
		\caption{Diagrammatic representation of the formula for the Local Green Marker (LGM) and the geometry of the problem. The black lines depict the borders of a unit supercell (see the main text). The blue region is the ``bulk'' of the system, separated from the borders by $l_x$ and $l_y$ sites (in the particular case depicted $l_x = l_y = 1$). The LGM at a given point $\bm{r}$ is a sum of terms represented by all possible loops originating at $\bm{r}$ and traversing the Euclidean space-time. In the thermodynamic limit, only loops fully contained in a single supercell - the green ones on the picture, give a non-vanishing contribution in the bulk. All other loops (depicted in red) have a vanishing contribution.}
		\label{SM:fig:supercell}
	\end{figure}
	
	In this supplemental material we present the detailed derivation of the Local Green Maker (LGM). Our starting point is a generalized Ishikawa-Matsuyama invariant proposed in \cite{zheng2019interaction}.  In a disordered system of size $n_x\times n_y$, the Hall conductivity may be approximated by tiling the system and treating it as a supercell in a larger superlattice. This construction is known as the Extended Infinite System (EIS) \cite{ceresoli2007orbital}. The position of a site in the EIS is denoted as $\bm{i} = \bm{R} + \bm{r}$, where $\bm{R}$ denotes a vector pointing to the origin of a supercell and $\bm{r}$ is a relative position in the unit cell:

	\begin{equation}
	\sigma_{xy} = \frac{\epsilon^{\mu\nu \rho}}{12 \pi N} \int_{-i\infty}^{i\infty} d\omega \;\sum_{\bm{\theta}}
	Tr\left(G^{\bm{\theta}}\partial_{\mu}\left[G^{\bm{\theta}}\right]^{-1}
	G^{\bm{\theta}}\partial_{\nu}\left[G^{\bm{\theta}}\right]^{-1}G^{\bm{\theta}}\partial_{\rho}\left[G^{\bm{\theta}}\right]^{-1}\right),
	\label{SM:eq:GIMP}
	\end{equation}	
	where $G^{\bm{\theta}}(\bm{r},\bm{r'})$ is the Fourier transform of the Matsubara Green's function with respect to the new effective periodicity, $\mu,\nu$ and $\rho$ take values $\{0,1,2\}$ corresponding to $\{ \omega, \bm{\theta}_x,\bm{\theta}_y\}$ and $N$ is the total number of sites in the EIS.
	
	The derivation consists of three main steps:

    \begin{itemize}
        \item First, we express $\sigma_{xy}$ as an average of a real space quantity $\mathfrak{G}(\bm{r})$. Then, we consider the Thermodynamic Limit (TL) $n_x, n_y \to \infty$. For simplicity, we consider the case $n_x = n_y$.  
        \item  Secondly, we prove that $\sigma_{xy}$ can be found by averaging $\mathfrak{G}(\bm{r})$ over the appropriately defined bulk of a supercell.
        \item  Finally, we demonstrate that for a large class of insulating interacting systems, the value of $\mathfrak{G}(\bm{r})$ in the bulk can be found using the Greens functions $G(\bm{r},\bm{r'})$ evaluated at points belonging to a single supercell. That is, for $\bm{r}$ in the bulk of a given supercell, contributions from other cells to $\mathfrak{G}(\bm{r})$ vanish.
    \end{itemize}

	 For the first step, we express the invariant (\ref{SM:eq:GIMP}) using quantities defined in real space-time, as opposed to Fourier space. It has been shown that the winding number of a function can be calculated from its Fourier transformed coefficients, provided that the function has a bounded mean oscillation \cite{brezis1995degree}. Therefore, we can safely go to real space and imaginary time in order to calculate the invariant (\ref{SM:eq:GIMP}):
		\begin{equation}
		\begin{split}
	G_{i\omega}^{\bm{\theta}}(\bm{r},\bm{r'}) &=  \sum_{\bm{R}'} G_{i\omega}(\bm{r},\bm{r'}+\bm{R}')e^{i\bm{\theta}(\bm{r} - \bm{r'} -\bm{R}')}\\
	\partial_{\mu}\left[G_{i\omega}^{\bm{\theta}}\right]^{-1}(\bm{r},\bm{r'}) &=  i \sum_{\bm{R}'}\left(\bm{r}^\mu - \bm{r'}^\mu -\bm{R}'^\mu\right) G^{-1}_{i\omega}(\bm{r},\bm{r'}+\bm{R}')e^{i\bm{\theta}(\bm{r} - \bm{r'} -\bm{R}')}\\
	G_{i\omega} &= \int_{0}^{\infty} d\tau G(0,\tau)e^{i\omega \tau},
	\label{SM:eq:inverseFourier}
	\end{split}
	\end{equation}
    where in the last line we have hidden the spatial variables to simplify the notation. We substitute these expressions into eqn.~(\ref{SM:eq:GIMP}) and unify the notation for space and imaginary time $\bm{r}^0\equiv\tau$, $\int_0^\beta d\tau\equiv\sum_{\bm{r}^0}$ . We denote the unit vector in the time direction as $\bm{e}^0$, allowing us obtain the following expression. Note that we, somewhat inconsistently, keep $\bm{R}$ and $\bm{\theta}$ to be $2d$ vectors:
    
    \begin{equation}
    \begin{split}
       \sigma_{xy} = -i \frac{\epsilon_{\mu\nu \rho}}{12 \pi N} \int_{-i\infty}^{i\infty} d\omega\sum_{\bm{\theta}} \sum_{\bm{R}_i,\bm{r}_i}  Tr\left( \vphantom{ \sum_{r_i} } G(\bm{r},\bm{r}_1+\bm{R}_1)\left[ (\bm{r}_2^{\mu} +\bm{R}_2^\mu-\bm{r}_1^{\mu} - \bm{R}_1^\mu)G^{-1}(\bm{r_1}+\bm{R}_1,\bm{r_2}+\bm{R}_2)\right]\right. \\ \times\left. G(\bm{r_2}+\bm{R}_2,\bm{r_3}+\bm{R}_3)\left[ (\bm{r}_4^{\nu} +\bm{R}_4^\nu-\bm{r}_3^{\nu} - \bm{R}_3^\nu)G^{-1}(\bm{r_3}+\bm{R}_3,\bm{r_4}+\bm{R}_4)\right] \right. \\ \times\left. G(\bm{r_4}+\bm{R}_4,\bm{r_5}+\bm{R}_5)  \left[(\bm{r}^{\rho}+\bm{R}_6^{\rho} -\bm{r}_5^{\rho} - \bm{R}_5^\rho) G^{-1}(\bm{r_5}+\bm{R}_5,\bm{r}+\bm{e}^0 \bm{r}_6^0 +\bm{R}_6)\right] \vphantom{ \sum_{r_i} }
       \right. \\ \times\left.  e^{i\bm{R}_6\bm{\theta}+i\omega \bm{r}_6^0} \right),  
    \end{split}
    \end{equation}
    where we have used translational invariance in space $G(\bm{r},\bm{r}' + \bm{R}' - \bm{R}) = G(\bm{r}+\bm{R}, \bm{r}' + \bm{R}')$ and imaginary time $G(0, \tau' - \tau)  = G(\tau,\tau')$. Evaluating the $\omega$ integration and $\bm{\theta}$ summation, we obtain:
    
    \begin{equation}
        \sigma_{xy} = \frac{1}{n_x n_y} \sum_{\bm{r}}\mathfrak{G}(\bm{r})
        \label{SM:eq:average}
    \end{equation}
    
     \begin{equation}
    \begin{split}
       \mathfrak{G}(\bm{r})=-i \frac{\epsilon_{\mu\nu \rho}}{6}    \sum_{\bm{R}_i,\bm{r}_i}Tr\left(\vphantom{ \sum_{r_i} } G(\bm{r},\bm{r}_1+\bm{R}_1)\left[ (\bm{r}_2^{\mu} +\bm{R}_2^\mu-\bm{r}_1^{\mu} - \bm{R}_1^\mu)G^{-1}(\bm{r_1}+\bm{R}_1,\bm{r_2}+\bm{R}_2)\right]\right. \\ \times\left. G(\bm{r_2}+\bm{R}_2,\bm{r_3}+\bm{R}_3)\left[ (\bm{r}_4^{\nu} +\bm{R}_4^\nu-\bm{r}_3^{\nu} - \bm{R}_3^\nu)G^{-1}(\bm{r_3}+\bm{R}_3,\bm{r_4}+\bm{R}_4)\right] \right. \\ \times\left. G(\bm{r_4}+\bm{R}_4,\bm{r_5}+\bm{R}_5)  \left[(\bm{r}^{\rho}-\bm{r}_5^{\rho} - \bm{R}_5^\rho) G^{-1}(\bm{r_5}+\bm{R}_5,\bm{r})\right] \vphantom{ \sum_{r_i} }\right).
   \label{SM:eq:LGMloops}
    \end{split}
    \end{equation}
    One can simplify eqn. \ref{SM:eq:LGMloops}, expanding the brackets and excluding the terms symmetric in ${\rho, \mu,\nu}$: 
    \begin{equation}
		\begin{split}
       \mathfrak{G}(\bm{r})=-i \frac{\epsilon_{\mu\nu \rho}}{6}  \sum_{\bm{r}_i} \sum_{\bm{R}_i} Tr \left( \vphantom{ \sum_{r_i} } G(\bm{r},\bm{r}_1 + \bm{R}_1)(\bm{r}_1^{\mu}+ \bm{R}_1^\mu) G^{-1}(\bm{r}_1+ \bm{R}_1,\bm{r}_2+ \bm{R}_2)(\bm{r}_2^\nu+ \bm{R}_2^\nu) \right. \\ \left. \times   G(\bm{r}_2,\bm{r}_3)(\bm{r}_3^\rho+ \bm{R}_3^\rho) G^{-1}(\bm{r}_3,\bm{r})\vphantom{ \sum_{r_i} } \right).
       \end{split}
       \label{SM:eq:LGM1}
  \end{equation}
    
    We can conclude that the winding number (\ref{SM:eq:GIMP}) can be presented diagrammatically (as demonstrated in fig. \ref{SM:fig:supercell}) as a sum over all possible ``loops'' consisting of products of Green's functions, their inverses, and ``vertices'' of $r$ variables components originating in a single supercell:
    
   Now, we separate the contributions in eqn. (\ref{SM:eq:average}) depending on whether they are from the bulk or the boundary of the supercell. We define bulk as the set of sites whose distance to the borders of the unit cell exceeds $l(n)$. Then, for the Hall conductivity we have:
   
   \begin{equation}
       \sigma_{xy} = \frac{1}{(n-2l)^2} \sum_{\bm{r}\in bulk}\mathfrak{G}(\bm{r}) + \frac{1}{4l(n-l)}\sum_{\bm{r}\notin bulk}\mathfrak{G}(\bm{r}).
       \label{SM:eq:split}
   \end{equation}
   
  As  we shall see below, $\mathfrak{G}(\bm{r}) = \mathcal O(1)$ in the limit $n\to\infty$. Therefore, the second term in the sum vanishes provided that $l$ as a function of $n$ grows slower than linearly with the system size.
  
  We proceed to the final step of the derivation, where we must prove that  $\mathfrak{G}(\bm{r})$ can be found from ``loops'' belonging to a single supercell of an insulating system, provided that interactions and hoppings decay exponentially and the limit $n\to\infty$ is taken. To this end, we separate the sum in eqn.~(\ref{SM:eq:LGMloops}) into two parts. The first part ($\mathfrak{G}_1$) consists of terms with all of the $\bm{R}_i$ equal to zero. This represents loops that stay within one supercell, shown with a green line in fig. \ref{SM:fig:supercell}. The second part ($\mathfrak{G}_2$) contains terms where at least two of the $\bm{R}_i$ are non-zero, representing loops that cross from one supercell to another, shown as red lines in fig. \ref{SM:fig:supercell}.
  
  \begin{equation}
       \mathfrak{G}(\bm{r})= \mathfrak{G}_1(\bm{r}) + \mathfrak{G}_2(\bm{r})
       \label{SM:eq:split}
   \end{equation}
  \begin{equation}
  \begin{split}
      \mathfrak{G}_1(\bm{r})=-i \frac{\epsilon_{\mu\nu \rho}}{6}  \sum_{\bm{r}_i} Tr \left( \vphantom{ \sum_{r_i} } G(\bm{r},\bm{r}_1)\bm{r}_1^{\mu} G^{-1}(\bm{r}_1,\bm{r}_2)\bm{r}_2^\nu    G(\bm{r}_2,\bm{r}_3)\bm{r}_3^\rho G^{-1}(\bm{r}_3,\bm{r})\vphantom{ \sum_{r_i} } \right)
       \end{split}
       \label{SM:eq:greenloops}
  \end{equation}
  \begin{equation}
  \begin{split}
    \mathfrak{G}_2(\bm{r})=-i \frac{\epsilon_{\mu\nu \rho}}{6}  \sum_{\bm{r}_i} \sum_{\{\bm{R}_i\}\in A} Tr \left( \vphantom{ \sum_{r_i} } G(\bm{r},\bm{r}_1 + \bm{R}_1)(\bm{r}_1^{\mu}+ \bm{R}_1^\mu) G^{-1}(\bm{r}_1+ \bm{R}_1,\bm{r}_2+ \bm{R}_2)(\bm{r}_2^\nu+ \bm{R}_2^\nu) \right. \\ \left. \times   G(\bm{r}_2,\bm{r}_3)(\bm{r}_3^\rho+ \bm{R}_3^\rho) G^{-1}(\bm{r}_3,\bm{r})\vphantom{ \sum_{r_i} } \right)
 \end{split} 
 \label{SM:eq:redloops}
  \end{equation}
 
For gapped systems with exponentially decaying interactions and hoppings, it can be proved that the Green's functions decay exponentially both in distance and time \cite{hastings2006spectral,nachtergaele2006lieb,wang2020tightening}. The same can be concluded for the inverse Green's functions, using the results of ref. \cite{jaffard1990proprietes,benzi2016localization}. Therefore $\exists\; \alpha > 0$ such that:
  
  \begin{equation}
\begin{split}
    G(\bm{r}+\bm{R}, \bm{r}'+\bm{R}') < A e^{- \alpha \norm{\bm{r}'+\bm{R}' - \bm{r}-\bm{R}}},\\
     G^{-1}(\bm{r}+\bm{R}, \bm{r}'+\bm{R}') < B e^{- \alpha \norm{\bm{r}'+\bm{R}' - \bm{r}-\bm{R}}}.
\end{split}
\label{SM:eq:GreenBounds}
  \end{equation}
  Inserting these bounds into eqn. \ref{SM:eq:greenloops}, one obtains:
  \begin{equation}
  \begin{split}
       \mathfrak{G}_1(\bm{r}) < C  \epsilon_{\mu\nu \rho} \sum_{\bm{r}_i}e^{-\alpha\left(\norm{\bm{r}_1-\bm{r}}+\norm{\bm{r}_2-\bm{r}_1}+\norm{\bm{r}_3-\bm{r}_2}+\norm{\bm{r}-\bm{r}_3}\right)}\bm{r}_1^{\mu} \bm{r}_2^{\nu} \bm{r}_3^{\rho} < \\ <   C  \epsilon_{\mu\nu \rho} \sum_{\bm{r}_i}e^{-\alpha\left(\norm{\bm{r}_1-\bm{r}}+\norm{\bm{r}_3-\bm{r}_2}+\norm{\bm{r}-\bm{r}_3}\right)}\bm{r}_1^{\mu} \bm{r}_2^{\nu} \bm{r}_3^{\rho},
 \end{split} 
 \label{SM:eq:boundgl}
  \end{equation}
  where in the second line we removed a non-negative term $\norm{\bm{r}_2-\bm{r}_1}$ in the exponent. This allows us to evaluate the sum over $\bm{r}_1$, using $\norm{r}> \frac{1}{2}\sum_\mu |\bm{r}_\mu| $, as it becomes a geometric series (in the time direction it is an integral over a decaying exponent in $\bm{r}^0$). Once the summation over $\bm{r}_1$ is done, the summation over $\bm{r}_2$ also simplifies to a geometric series. Proceeding further, one therefore can conclude that $\mathfrak{G}_1 = \mathcal O(1)$ in the thermodynamic limit.

  As for eqn. (\ref{SM:eq:greenloops}), one can obtain the following inequality:
   \begin{equation}
  \begin{split}
       \mathfrak{G}_2(\bm{r}) < \tilde{C}  \epsilon_{\mu\nu \rho} \sum_{\{R_i\}\in A}\sum_{\bm{r}_i}e^{-\alpha (\norm{\bm{r}_1+\bm{R}_1-\bm{r}} +
       \norm{\bm{r_2}+\bm{R}_2-\bm{r_1}-\bm{R}_1}+\norm{\bm{r_3}+\bm{R}_3-\bm{r_2}-\bm{R}_2}+\norm{\bm{r}-\bm{r_3}-\bm{R}_3})} \\\times  (\bm{r}_1^{\mu} +\bm{R}_1^\mu)  (\bm{r}_2^{\nu} +\bm{R}_2^\nu)(\bm{r}_3^\rho +\bm{R}_3^{\rho}).
 \end{split} 
 \label{SM:eq:boundrl}
  \end{equation}
  
  It is sufficient to prove that the sum in eqn.~(\ref{SM:eq:boundrl}) vanishes when $\bm{R}_1\ne0$. To show this, one can notice that:
  
  \begin{equation}
   \sum_{R_1\ne0,\bm{r_1}}e^{-\alpha (\norm{\bm{r}_1+\bm{R}_1-\bm{r}} )} < e^{-l(n)\alpha}\sum_{R_1}e^{-\alpha \norm{\bm{R}_1} },
  \end{equation}
  
 and then evaluate the geometric series using the same tactic as that used for  $\mathfrak{G}_1$, to see that $\mathfrak{G}_2(\bm{r})= \mathcal O (\exp(-l(n)\alpha))$ which vanishes in the thermodynamic limit under the choice $l(n) = \log(n)$. $\blacksquare$

  \end{widetext}

\end{document}